\documentclass[showpacs,amssymb,twocolumn,aps,nofootinbib]{revtex4}
\usepackage{amsmath}
\usepackage{amstext}
\usepackage{amsopn}
\usepackage{amsfonts}
\usepackage{amssymb}
\usepackage{amssymb}
\usepackage{bbm}
\usepackage{accents}
\usepackage{empheq}%for overbracket
\usepackage{graphicx}
\usepackage{epsf}
\usepackage{graphics}
\usepackage{xcolor}
\usepackage[latin1]{inputenc}

\begin{document}

\title{Bogoliubov transformation and the thermal operator representation in the real time formalism}

\author{Ashok Das$^{a,b}$, Atri Deshamukhya$^{c}$ Pushpa Kalauni$^{d}$ and S. Panda$^{b,e,f}$}
\affiliation{$^a$Department of Physics and Astronomy, University of Rochester,
Rochester, NY 14627-0171, USA}
\affiliation{$^b$ Institute of Physics, Sachivalaya Marg, Bhubaneswar 751005, India}
\affiliation{$^c$Department of Physics, Assam University, Silchar 788011, India}
\affiliation{$^d$Homer L. Dodge Department of Physics and Astronomy, University
of Oklahoma, Norman, OK 73019, USA}
\affiliation{$^e$National Institute of Science Education and Research, Jatni, Bhubaneswar 752050, India}
\affiliation{$^f$Homi Bhabha National Institute, Anushakti Nagar, Mumbai 400085, India}

\begin{abstract}
It has been shown earlier \cite{brandt,brandt1} that, in the mixed space, there is an unexpected simple relation  between any finite temperature graph and its zero temperature counterpart through a multiplicative scalar operator (termed thermal operator) which carries the entire temperature dependence. This was shown to hold only in the imaginary time formalism and the closed time path ($\sigma=0$) of the real time formalism (as well as for its conjugate $\sigma=1$). We study the origin of this operator from the more fundamental Bogoliubov transformation which acts, in the momentum space, on the doubled space of fields in the real time formalisms \cite{takahashi,umezawa,pushpa}. We show how the ($2\times 2$) Bogoliubov transformation matrix naturally leads to the scalar thermal operator for $\sigma=0,1$ while it fails for any other value  $0<\sigma<1$. This analysis also suggests that a generalized scalar thermal operator description, in the mixed space, is possible even for $0<\sigma<1$. We also show the existence of a scalar thermal operator relation in the momentum space.
\end{abstract}
\maketitle

\section{Introduction}

\begin{figure}[h]
\includegraphics[scale=0.3]{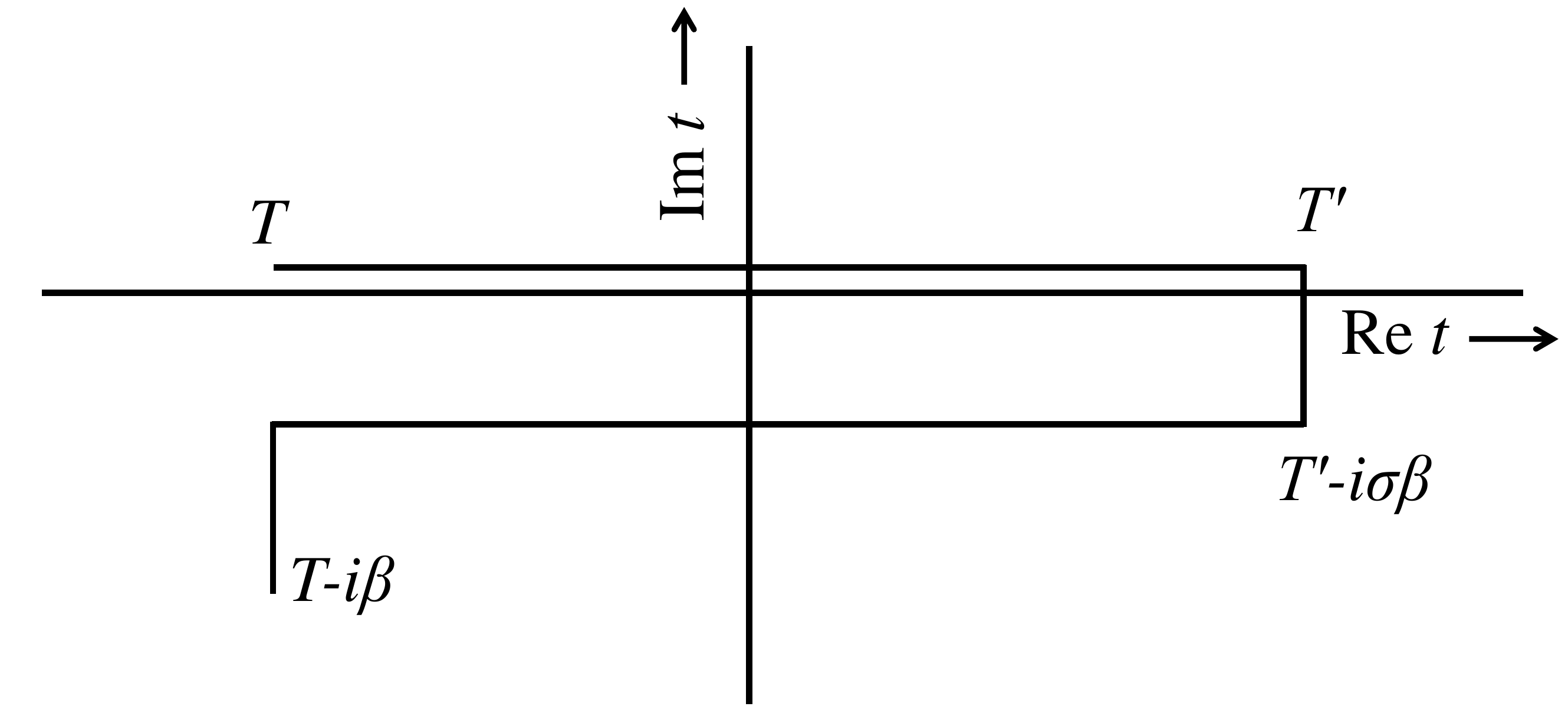}
\caption{The general time path contour in the complex $t$-plane with $0\leq\sigma\leq 1$ where $\beta$ denotes the inverse temperature in units of the Boltzmann constant. In the real time formalism, $T$ can be thought of as a time in the infinite past ($T\rightarrow -\infty$), while $T'$ denotes a time in the infinite future ($T\rightarrow \infty$). In this description, imaginary time formalism corresponds to $T=T'=0$. (Please note that in the text of the paper, on the other hand, $T$ represents the equilibrium temperature.)}
\label{f1}
\end{figure}

Equilibrium thermal field theory can be equally well described in the imaginary time formalism (Matsubara formalism) \cite{matsubara} or in the real time formalism \cite{bellac,das}. In the real time formalism, there is a one parameter family of paths in the complex $t$-plane (see Fig. \ref{f1}) \cite{matsumoto,nakano} which give the same result for the thermal ensemble averages of physical observables. In particular, for $\sigma=0$, the real time description is known as the closed time path formalism \cite{schwinger,bakshi,keldysh} while $\sigma=\frac{1}{2}$ leads to thermofield dynamics \cite{takahashi,umezawa,umezawa1}. 

In the real time formalism, the degrees of freedom are doubled because of the two real branches of the time contour in the complex $t$-plane (there is no doubling in the imaginary time formalism). Correspondingly, the Hilbert space for the system, in a real time description, is a product space of two Hilbert spaces - one for that of the physical system and the other can be thought of as that of the heat bath (environment). The real time formalism (for any value of the arbitrary parameter $\sigma$) allows for a path integral description \cite{bellac,das} as well as an operator description \cite{pushpa}. Thermofield dynamics ($\sigma=\frac{1}{2}$) has an operator description with the usual Dirac inner product for the doubled Hilbert space \cite{takahashi,umezawa,khanna}. For any other value of $\sigma$, however, the Hilbert space develops a modified inner product (which depends on the value of $\sigma$) \cite{pushpa, greenwood,greenwood1}. 

A thermal vacuum as well as a thermal Hilbert space can be defined (in the real time formalism) from the doublet of  fields (thermal doublet) through a Bogoliubov transformation \cite{takahashi,umezawa,pushpa}. The doublet of thermal field operators can then be related to the original (zero temperature) doublet of field operators (that of the original field and the doubled field for the ``heat bath") through a $2\times 2$ Bogoliubov transformation matrix in the momentum space. Correspondingly, the Feynman propagator in momentum space, say for a real scalar field theory on the general path (Fig. \ref{f1}), becomes a $2\times 2$ matrix propagator related to the zero temperature propagator through the $2\times 2$ Bogoliubov transformation matrix \cite{pushpa},
\begin{equation}
G^{(\sigma,T)} (p) = U^{(\sigma)}(T, p) G^{(\sigma,T=0)} (p) (U^{(\sigma)})^{\rm T} (T, -p),\label{BT}
\end{equation}
where $(U^{(\sigma)})^{\rm T}$ denotes the matrix transpose of $U^{(\sigma)}$ and $T$ represents the equilibrium temperature.

Since vertices, in a thermal field theory, do not carry any temperature dependence, all the temperature dependence, in a Feynman diagram at finite temperature, is carried by the internal propagators. However, as we have noted in \eqref{BT}, the $2\times 2$ thermal matrix propagator, in the momentum space, is related to the zero temperature propagator through the ($2\times 2$) Bogoliubov transformation matrix depending on temperature. Therefore, there is a natural factorization of any thermal Feynman graph, in momentum space, in terms of the zero temperature graph and a product of $2\times 2$ Bogoliubov transformation matrices carrying the temperature dependence (and a factor of $\sigma_{3}$ coming from every internal thermal vertex \cite{das}). However, keeping track of the matrix indices in a general graph is not so easy and, therefore, such a factorization, even though it arises naturally, is not very useful in practical calculations.

On the other hand, it has been shown \cite{brandt,brandt1}, both in the imaginary time formalism and the closed time path formalism of real time, that each finite temperature Feynman diagram in a quantum field theory, in the mixed space $(t,\mathbf{p})$, is related to the corresponding zero temperature diagram through a product of simple multiplicative scalar operators which carries the entire temperature dependence. For example, the thermal propagator (for a real scalar field theory) in the closed time path in the mixed space is related to the zero temperature propagator simply as
\begin{align}
G_{CT}^{(T)} (t, \mathbf{p}) & = {\cal O}^{(T)}(E) G_{CT}^{(T=0)} (t, \mathbf{p})\notag\\
&= \left(1 + n_{B}(E) (1 - S(E))\right) G_{CT}^{(T=0)} (t, \mathbf{p}),\label{TO}
\end{align} 
where $T$ denotes the equilibrium temperature, 
\begin{equation}
E= (\mathbf{p}^{2}+m^{2})^{\frac{1}{2}}, \label{E}
\end{equation}
$n_{B}(E)$ represents the Bose-Einstein distribution and $S(E)$ is a reflection operator which changes $E\rightarrow -E$ in any function, namely,
\begin{equation}
S(E)f(E) = f(-E).\label{reflection}
\end{equation}
(We would like to comment here parenthetically that there is no doubling of fields in the imaginary time formalism and, consequently, there is a natural factorization of the propagator in terms of a scalar thermal operator in the mixed space. The imaginary time formalism, on the other hand, has other subtleties involving the difference in the range of time integration at finite temperature and at zero temperature which have been discussed in detail in \cite{brandt}.) This simple  factorization of diagrams into a temperature dependent part and a zero temperature part is quite interesting and useful  since the Feynman diagram calculations at non-zero temperature are, in general, tedious compared to zero temperature. The thermal operator description of finite temperature diagrams has also led to a better understanding of several other questions at finite temperature \cite{niegawa,brandt2,brandt3,brandt4,frenkel,frenkel1}. On the other hand, it is also known \cite{brandt} that such a factorization, in terms of a simple scalar thermal operator, does not hold for a general time contour (see Fig. \ref{f1}) for which $\sigma\neq 0,1$.

In this paper, we would like to understand how the scalar thermal operator ${\cal O}^{(T)}(E)$, in the mixed space, arises from the more fundamental Bogoliubov transformation matrix $U(T,p)$ (in momentum space) in the real time formalisms for $\sigma=0,1$ and why it fails for other values of $\sigma$. Furthermore, we would like to study whether such an analysis may suggest a possible generalization of the thermal operator representation which holds for all values of $0\leq \sigma\leq 1$. In section {\bf II}, we recapitulate briefly various properties of the thermal propagator on a general path both in the momentum space as well as in the mixed space. This suggests that the two cases $\sigma=0,1$ and $0<\sigma<1$ are inherently different and need to be studied separately. The difference arises mainly because when $\sigma=0,1$, all four matrix elements of the zero temperature propagator are nontrivial while for other values of $\sigma$, the zero temperature propagator is a diagonal matrix.  In section {\bf III}, we discuss the Bogoliubov transformation matrix (in momentum space) for the closed time path formalism, $\sigma=0$ (the other case, $\sigma=1$ is related by a symmetry $\sigma\rightarrow 1-\sigma$). We show that the Bogoliubov transformation, in momentum space, leads to a matrix factorization of the propagator in the mixed space as well. From the properties of this factorizing matrix, we show how a scalar thermal operator arises for $\sigma=0,1$. In section {\bf IV}, we study the other case for $0<\sigma<1$ in detail. From the Bogoliubov transformation matrix in the momentum space, we show that the thermal propagator naturally factorizes in the mixed space as well. However, a direct calculation shows that the factorizing matrix does not lead to a scalar thermal operator in this case. On the other hand, studying the properties of this factorizing matrix, we show that a scalar thermal operator is possible if we treat the $2\times2$ matrix propagator at zero temperature in a limiting manner (containing all four elements and not just the diagonal elements). In section {\bf V}, we summarize our results and, in the appendix, we show how the explicit matrix factorization in mixed space arises from the more fundamental Bogoliubov transformation matrix in the momentum space. For the interested reader, we point out that our derivation of the scalar thermal operator for the closed time path is given in \eqref{scalar1}-\eqref{TO1}, while for the case $0<\sigma<1$, the final result is given in \eqref{TOR} where the usual zero temperature propagator is replaced by a generalized $\overline{G}^{(\sigma)}(t,\mathbf{p})$ which is given in \eqref{Gbar}. 

\section{Thermal propagator for an arbitrary path}

The thermal propagator, $G^{(\sigma,T)}$, of a real scalar field for a general path in the real time formalism (see Fig. \ref{f1}) is a $2\times 2$ matrix
\begin{equation}
G^{(\sigma,T)} = \begin{pmatrix}
G^{(\sigma,T)}_{11} & G^{(\sigma,T)}_{12}\\
\noalign{\vskip 2mm}%
G^{(\sigma,T)}_{21} & G^{(\sigma,T)}_{22}
\end{pmatrix},
\end{equation}
whose components in the momentum space at finite temperature are given by ($0\leq\sigma\leq 1$, $T$ represents temperature and $\beta$ denotes inverse temperature in units of the Boltzmann constant)
\begin{align}
G_{11}^{(\sigma, T)} (p) & = \frac{i}{p^{2}-m^{2}+i\epsilon} + 2\pi n_{B}(|p_{0}|)\delta(p^{2}-m^{2}),\notag\\
G_{12}^{(\sigma,T)} (p) & = 2\pi e^{\sigma\beta p_{0}} (\theta(-p_{0}) + n_{B}(|p_{0}|))\delta (p^{2}-m^{2}),\notag\\
G_{21}^{(\sigma,T)} (p) & = 2\pi e^{-\sigma\beta p_{0}}(\theta(p_{0}) + n_{B} (|p_{0}|))\delta (p^{2}-m^{2}),\notag\\
G_{22}^{(\sigma,T)} (p) & = -\frac{i}{p^{2}-m^{2}-i\epsilon} + 2\pi n_{B} (|p_{0}|) \delta(p^{2}-m^{2}).\label{generalpath}
\end{align}

In the mixed space, the propagator defined as
\begin{equation}
G^{(\sigma, T)} (t,\mathbf{p}) = \int \frac{dp_{0}}{2\pi}\, e^{-ip_{0}t}\, G^{(\sigma, T)} (p),
\end{equation}
has the components
\begin{align}
G_{11}^{(\sigma,T)} (t,\mathbf{p}) & = \frac{1}{2E}\big((\theta(t) + n_{B}(E))e^{-iEt}\notag\\
&\qquad\qquad + (\theta(-t)+n_{B}(E))e^{iEt}\big),\notag\\
G_{12}^{(\sigma,T)}(t,\mathbf{p}) & = \frac{1}{2E}\big(n_{B}(E) e^{-iE(t+i\sigma\beta)}\notag\\
&\qquad\qquad + (1+n_{B}(E))e^{iE(t+i\sigma\beta)}\big),\notag\\
G_{21}^{(\sigma,T)}(t,\mathbf{p}) & = \frac{1}{2E}\big((1+n_{B}(E))e^{-iE(t-i\sigma\beta)}\notag\\
&\qquad\qquad+n_{B}(E)e^{iE(t-i\sigma\beta)}\big),\notag\\
G_{22}^{(\sigma,T)}(t,\mathbf{p}) & = \frac{1}{2E}\big((\theta(-t)+n_{B}(E))e^{-iEt}\notag\\
&\qquad\qquad + (\theta(t)+n_{B}(E)) e^{iEt}\big),\label{generalpathmixed}
\end{align}
where, as we have pointed out earlier, $E=(\mathbf{p}^{2}+m^{2})^{\frac{1}{2}}$ and $n_{B}(E)$ denotes the Bose-Einstein distribution. 

There are several things to note from the structures of the components of the propagator in \eqref{generalpath} and \eqref{generalpathmixed}. First, the diagonal components of the propagator are independent of the arbitrary parameter $\sigma$ which characterizes the path in Fig. \ref{f1}. Second, the off-diagonal elements $G_{12}^{(\sigma,T)}$ and $G_{21}^{(\sigma,T)}$, which do depend on $\sigma$, are related simply as (both in the momentum space as well as in the mixed space)
\begin{equation}
G_{12}^{(\sigma,T)} = G_{21}^{(1-\sigma,T)},\quad G_{21}^{(\sigma,T)} = G_{12}^{(1-\sigma,T)},
\end{equation}
so that it is sufficient to study the propagator in the parameter range $0\leq \sigma\leq \frac{1}{2}$. However, for completeness we will allow the parameter $\sigma$ to take on the full range of values $0\leq\sigma\leq 1$ (keeping in mind this symmetry). Finally, we note that the off-diagonal components in \eqref{generalpath} and \eqref{generalpathmixed} vanish for $T=0$ ($\beta\rightarrow \infty$) when $\sigma\neq 0, 1$. Therefore, the zero temperature propagator for closed time path ($\sigma=0$), for example, has four nontrivial components (the same is true for $\sigma=1$) while, for any value $0<\sigma< 1$, the only nontrivial components of the zero temperature propagator are the two diagonal elements. As a result, the analysis for the two cases, $\sigma=0,1$ and for $0<\sigma< 1$, needs to be done separately. In the next section, we will study the propagator for the closed time path corresponding to $\sigma=0$ with comments on the case $\sigma=1$.

\section{Scalar thermal operator from Bogoliubov transformation for closed time path}

The momentum space propagator in the closed time path is denoted by the $2\times 2$ matrix  (in the closed time path, the components are conventionally labelled by $\pm$)
\begin{equation}
G^{(T)}_{CT} (p) = \begin{pmatrix}
G^{(T)}_{++}(p) & G^{(T)}_{+-} (p)\\
\noalign{\vskip 2mm}%
G^{(T)}_{-+} (p) & G^{(T)}_{--} (p)
\end{pmatrix},
\end{equation}
whose components at finite temperature are given by (see \eqref{generalpath} with $\sigma=0$)
\begin{align}
G_{++}^{(T)} (p) & = \frac{i}{p^{2}-m^{2}+i\epsilon} + 2\pi n_{B}(|p_{0}|)\delta(p^{2}-m^{2}),\notag\\
G_{+-}^{(T)} (p) & = 2\pi (\theta(-p_{0}) + n_{B}(|p_{0}|))\delta (p^{2}-m^{2}),\notag\\
G_{-+}^{(T)} (p) & = 2\pi (\theta(p_{0}) + n_{B} (|p_{0}|))\delta (p^{2}-m^{2}),\notag\\
G_{--}^{(T)} (p) & = -\frac{i}{p^{2}-m^{2}-i\epsilon} + 2\pi n_{B} (|p_{0}|) \delta(p^{2}-m^{2}).\label{CT-T}
\end{align}
Correspondingly, the zero temperature propagator, in the closed time path has the components
\begin{align}
G_{++}^{(T=0)} (p) & = \frac{i}{p^{2}-m^{2}+i\epsilon},\notag\\
G_{+-}^{(T=0)} (p) & = 2\pi \theta(-p_{0})\delta (p^{2}-m^{2}),\notag\\
G_{-+}^{(T=0)} (p) & = 2\pi \theta(p_{0})\delta (p^{2}-m^{2}),\notag\\
G_{--}^{(T=0)} (p) & = -\frac{i}{p^{2}-m^{2}-i\epsilon}.\label{CT-zero}
\end{align}
As alluded to earlier, the zero temperature propagator, in the closed time path, has four nontrivial components just like at finite temperature (the same is true for $\sigma=1$ as well). This is not the case for $0<\sigma< 1$ as we will see in the next section.

The Bogoliubov transformation matrix, relating the finite temperature propagator in \eqref{CT-T} to the zero temperature propagator in \eqref{CT-zero} in the closed time path, has the form
\begin{align}
U_{CT} (T,p) & = \frac{1}{(2\sinh \beta |p_{0}|)^{\frac{1}{2}}} \begin{pmatrix}
e^{\frac{\beta|p_{0}|}{2}} & e^{-\frac{\beta|p_{0}|}{2}}\\
e^{-\frac{\beta|p_{0}|}{2}} & e^{\frac{\beta|p_{0}|}{2}}
\end{pmatrix}\notag\\
& = U_{CT}(T,-p) = U_{CT}^{\rm T} (T,-p),\label{pCT-BTmatrix}
\end{align}
so that we can write (see \eqref{BT})
\begin{equation}
G_{CT}^{(T)} (p) = U_{CT}(T,p) G_{CT}^{(T=0)}(p) U_{CT} (T,p).\label{BT-CTp}
\end{equation}
In this case, the Bogoliubov transformation matrix satisfies $U^{\dagger} = U^{\rm T} = U$. Furthermore, it can be checked to satisfy $U^{\dagger} \sigma_{3}U = \sigma_{3}$ (where $\sigma_{3}$ is the Pauli matrix) reflecting the fact that the Bogoliubov transformation belongs to the noncompact group $SO(2,1)$.

In the mixed space, the finite temperature propagator in the closed time path ($\sigma=0$) has the components (see \eqref{generalpathmixed})
\begin{align}
G_{++}^{(T)}(t,\mathbf{p}) & = \frac{1}{2E}\big((\theta(t)+n_{B}(E)) e^{-iEt}\notag\\
& \qquad\qquad + (\theta(-t)+n_{B}(E))e^{iEt}\big),\notag\\
G_{+-}^{(T)}(t,\mathbf{p}) & = \frac{1}{2E}\big(n_{B}(E) e^{-iEt}\notag\\
&\qquad\qquad + (1+n_{B}(E))e^{iEt}\big),\notag\\
G_{-+}^{(T)}(t,\mathbf{p}) & = \frac{1}{2E}\big((1+n_{B}(E))e^{-iEt}\notag\\
&\qquad\qquad+n_{B}(E)e^{iEt}\big),\notag\\
G_{--}^{(T)}(t,\mathbf{p}) & = \frac{1}{2E}\big((\theta(-t)+n_{B}(E))e^{-iEt}\notag\\
&\qquad\qquad + (\theta(t)+n_{B}(E)) e^{iEt}\big),\label{mixedCT-T}
\end{align}
whereas at zero temperature, the components have the forms
\begin{align}
G_{++}^{(T=0)} (t,\mathbf{p}) & = \frac{1}{2E}\big(\theta(t)e^{-iEt} + \theta(-t)e^{iEt}\big),\notag\\
G_{+-}^{(T=0)} (t,\mathbf{p}) & = \frac{1}{2E}\,e^{iEt},\notag\\
G_{-+}^{(T=0)} (t,\mathbf{p}) & = \frac{1}{2E}\, e^{-iEt},\notag\\
G_{--}^{(T=0)} (t,\mathbf{p}) & = \frac{1}{2E}\big(\theta(-t)e^{-iEt} + \theta(t)e^{iEt}\big).\label{mixedCT-zero}
\end{align}

The finite temperature propagator in \eqref{mixedCT-T} can now be written in a factorized form in terms of the zero temperature propagator in \eqref{mixedCT-zero} as
\begin{align}
G_{CT}^{(T)} (t,\mathbf{p}) & = \overline{U} (T,\mathbf{p}) G_{CT}^{(T=0)} (t,\mathbf{p}) \overline{U}^{\rm T}(T,-\mathbf{p})\notag\\
& = \overline{U} (T,\mathbf{p}) G_{CT}^{(T=0)} (t,\mathbf{p}) \overline{U}(T,\mathbf{p}),\label{matrixfactorization}
\end{align}
where
\begin{align}
\overline{U}(T,\mathbf{p}) & = \frac{1}{(2\sinh\beta E)^{\frac{1}{2}}}\begin{pmatrix}
e^{\frac{\beta E}{2}} & e^{-\frac{\beta E}{2}}\\
e^{-\frac{\beta E}{2}} & e^{\frac{\beta E}{2}}
\end{pmatrix}\nonumber\\
& = \overline{U}(T,-\mathbf{p}) = \overline{U}^{\rm T} (T,-\mathbf{p}),\label{mixedCT-BTmatrix}
\end{align}
with $E$ defined in \eqref{E}. Since the diagonal elements of the zero temperature propagator in \eqref{CT-zero} have poles at $p_{0}= \pm E$ where the off-diagonal elements also have nontrivial contributions and since $U_{CT}(T,p)$ depends only on the magnitude $|p_{0}|$, it is clear that the matrix $\overline{U}$ in \eqref{mixedCT-BTmatrix} arises naturally from the momentum space Bogoliubov transformation matrix \eqref{pCT-BTmatrix} when we Fourier transform \eqref{BT-CTp} to the mixed space.  We note that the factorizing matrix, $\overline{U}(T,\mathbf{p})$, is independent of the $t$ coordinate in this case, which we will see in the next section not to be the case for $0<\sigma<1$. The factorizability of the thermal propagator in the mixed space is a direct consequence of the Bogoliubov transformation relating the finite temperature and zero temperature propagators in the momentum space. As in the momentum space, we note from \eqref{mixedCT-BTmatrix} that 
\begin{equation}
\overline{U}(T,\mathbf{p}) = \overline{U}^{\dagger} (T,\mathbf{p}) = \overline{U}^{\rm T} (T,\mathbf{p}),\label{Ubarprop}
\end{equation} 
and that $\overline{U}$ satisfies the group property
\begin{equation}
\overline{U}^{\dagger} (T,\mathbf{p}) \sigma_{3} \overline{U}(T,\mathbf{p}) = \sigma_{3}.\label{overlineUgroup}
\end{equation}

\subsection{Derivation of the scalar thermal operator}

As we have already pointed out, a matrix factorization of the thermal propagator is always possible. In this subsection we will discuss how a scalar thermal operator (see \eqref{TO}) arises from the matrix factorization in the case of the closed time path. We note that we can write the matrix factorization in \eqref{matrixfactorization} as (we are suppressing the arguments for simplicity)
\begin{align}
G_{CT}^{(T)} & = \overline{U}G_{CT}^{(T=0)}\overline{U} = (\overline{U}\sigma_{3})(\sigma_{3}G_{CT}^{(T=0)})\overline{U}\notag\\
&= (\overline{U}\sigma_{3})\big(\overline{U}(\sigma_{3}G_{CT}^{(T=0)}) + [\sigma_{3}G_{CT}^{(T=0)},\overline{U}]\big)\notag\\
& = G_{CT}^{(T=0)} + \overline{U}\sigma_{3} [\sigma_{3}G_{CT}^{(T=0)},\overline{U}],\label{scalar1}
\end{align}
where we have used \eqref{Ubarprop} and \eqref{overlineUgroup} in the last step. The zero temperature part of the propagator comes out naturally in \eqref{scalar1} and, therefore, the second term must correspond to the finite temperature correction to the propagator.

To proceed further, let us note some identities following from \eqref{mixedCT-zero},
\begin{align}
G_{++}^{(T=0)} + G_{--}^{(T=0)} & = G_{+-}^{(T=0)}+G_{-+}^{(T=0)},\notag\\
S(E) G_{\pm\pm}^{(T=0)} & = - G_{\mp\mp}^{(T=0)},\notag\\
S(E) G_{\pm\mp}^{(T=0)} & = - G_{\mp\pm}^{(T=0)}.\label{identities}
\end{align}
Using these, it can be shown that
\begin{align}
\lefteqn{\overline{U}\sigma_{3}[\sigma_{3}G_{CT}^{(T=0)},\overline{U}]}\notag\\
& = n_{B}(E)\begin{pmatrix}
G_{++}^{(T=0)}+G_{--}^{(T=0)} & G_{+-}^{(T=0)}+G_{-+}^{(T=0)}\\
\noalign{\vskip 2pt}%
G_{-+}^{(T=0)}+G_{+-}^{(T=0)} & G_{--}^{(T=0)}+G_{++}^{(T=0)}
\end{pmatrix}\notag\\
& = n_{B}(E) (1 - S(E)) \begin{pmatrix}
G_{++}^{(T=0)} & G_{+-}^{(T=0)}\\
\noalign{\vskip 2pt}%
G_{-+}^{(T=0)} & G_{--}^{(T=0)}
\end{pmatrix}\notag\\
& = n_{B}(E) (1-S(E)) G_{CT}^{(T=0)},
\end{align}
where $S(E)$ denotes the reflection operator defined in \eqref{reflection}. As a result, we see from \eqref{scalar1} that we can write the finite temperature propagator in mixed space as
\begin{align}
G_{CT}^{(T)} (t,\mathbf{p}) & = (1 + n_{B}(E) (1-S(E)))G_{CT}^{(T=0)}(t,\mathbf{p})\notag\\
& = {\cal O}^{(T)}(E) G_{CT}^{(T=0)} (t,\mathbf{p}),\label{TO1}
\end{align}
where ${\cal O}^{(T)}$ coincides with the scalar thermal operator defined in \eqref{TO}. A completely parallel analysis can be done for the case $\sigma=1$ and leads to the fact that the scalar thermal operator \eqref{TO} naturally arises from the corresponding Bogoliubov transformation matrix. 

This analysis shows how the scalar thermal operator naturally arises, in the mixed space, from the momentum space Bogoliubov transformation matrix for closed time path (as well as for $\sigma=1$). In the next section, we will do the corresponding analysis for a general path for which $\sigma\neq 0,1$.

\section{Bogoliubov transformation and the scalar thermal operator for a general path}

At finite temperature, the propagator for $0\leq\sigma\leq1$ has four nontrivial components. For $\sigma=0,1$ this is also the case at zero temperature (see, for example, \eqref{CT-zero} and \eqref{mixedCT-zero}). For $\sigma\neq 0,1$ (namely, when $0<\sigma<1$), on the other hand, the zero temperature propagator becomes diagonal, the diagonal elements coinciding with those of the zero temperature propagator of the closed time path \eqref{CT-zero} and \eqref{mixedCT-zero}. For example, in the momentum space the components of the zero temperature propagator  have the forms
\begin{align}
G_{11}^{(\sigma, T=0)} (p) & = \frac{i}{p^{2}-m^{2}+i\epsilon},\notag\\
G_{12}^{(\sigma, T=0)}(p) & = 0,\notag\\
G_{21}^{(\sigma,T=0)} (p) & = 0,\notag\\
G_{22}^{(\sigma, T=0)} (p) & = - \frac{i}{p^{2}-m^{2}-i\epsilon},\label{general-zero}
\end{align}
while in the mixed space, they are given by
\begin{align}
G_{11}^{(\sigma, T=0)} (t,\mathbf{p}) & = \frac{1}{2E}\big(\theta(t)e^{-iEt} + \theta(-t)e^{iEt}\big),\notag\\
G_{12}^{(\sigma,T=0)} (t,\mathbf{p}) & = 0,\notag\\
G_{21}^{(\sigma,T=0)} (t,\mathbf{p}) & = 0,\notag\\
G_{22}^{(\sigma,T=0)} (t,\mathbf{p}) & = \frac{1}{2E}\big(\theta(-t)e^{-iEt} + \theta(t)e^{iEt}\big).\label{mixedgeneral-zero}
\end{align}
As a result, it is clear that a (well behaved) scalar thermal operator acting on the trivial off-diagonal components in \eqref{mixedgeneral-zero} cannot generate the off-diagonal terms in the thermal propagator in \eqref{generalpathmixed}.

On the other hand, a matrix factorization of the thermal propagator in terms of the zero temperature propagator (even if it is diagonal) is always possible through the Bogoliubov transformation matrix. We would like to discuss next how the momentum space Bogoliubov transformation matrix fails to lead to a scalar thermal operator in the mixed space in this case and whether an analysis from the point of view of the Bogoliubov transformation matrix can suggest a possible way out for a scalar thermal operator in the case of a general path $0<\sigma<1$. 

It is known \cite{pushpa} that, for a general path, the finite temperature and the zero temperature propagators (\eqref{generalpath} and \eqref{mixedgeneral-zero} respectively)  are related by
\begin{equation}
G^{(\sigma,T)} (p) = U^{(\sigma)}(T,p) G^{(\sigma,T=0)}(p)(U^{(\sigma)})^{\rm T} (T,-p),\label{BT-generalp}
\end{equation}
where the $2\times 2$ Bogoliubov transformation matrix has the form
\begin{equation}
U^{(\sigma)} (T,p) = (n_{B}(|p_{0}|))^{\frac{1}{2}}\begin{pmatrix}
e^{\frac{\beta|p_{0}|}{2}} & e^{(\sigma-\frac{1}{2})\beta p_{0}}\\
e^{-(\sigma-\frac{1}{2})\beta p_{0}} & e^{\frac{\beta|p_{0}|}{2}}
\end{pmatrix}.\label{pgeneral-BTmatrix}
\end{equation}
It follows from \eqref{pgeneral-BTmatrix} that
\begin{align}
(U^{(\sigma)})^{\dagger}(T,p) & = (U^{(\sigma)})^{\rm T} (T,p)\notag\\
& = U^{(\sigma)} (T,-p) = U^{(1-\sigma)} (T,p).\label{identities1}
\end{align}
As a result, the thermal propagator \eqref{BT-generalp} can also be written as
\begin{equation}
G^{(\sigma,T)} (p) = U^{(\sigma)}(T,p) G^{(\sigma,T=0)}(p)U^{(\sigma)} (T,p).\label{BT-generalp1}
\end{equation}

The $SO(2,1)$ group property of the Bogoliubov transformation, in the present case, is given by
\begin{equation}
(U^{(1-\sigma)})^{\dagger}(T,p)\sigma_{3}U^{(\sigma)} (T,p) = \sigma_{3}.
\end{equation}
This unusual relation of the group property, in this case, is a consequence of the fact that the thermal Hilbert space of states develops a nonstandard inner product for nontrivial $\sigma$ (for $\sigma\neq \frac{1}{2}$) which leads to a modified definition of the adjoint of an operator. (We refer the readers to \cite{pushpa} for further details on this.)

We note from \eqref{pgeneral-BTmatrix} that the Bogoliubov transformation matrix, in this case, is not simply a function of $|p_{0}|$, unlike the Bogoliubov transformation matrix for closed time path in \eqref{pCT-BTmatrix}, but depends on $p_{0}$ as well. As a result, the Fourier transform of \eqref{BT-generalp1} to the mixed space needs to be done carefully (which we do in some detail in the appendix). The poles of the zero temperature propagator \eqref{general-zero} are still at $p_{0}=\pm E = \pm (\mathbf{p}^{2}+m^{2})^{\frac{1}{2}}$. However, the off-diagonal elements of the Bogoliubov transformation matrix in \eqref{pgeneral-BTmatrix} are now sensitive to the choice of the ($p_{0}$) contour of integration. The Fourier transform of \eqref{BT-generalp1} to mixed space leads to a matrix factorization of the thermal propagator in the mixed space of the form
\begin{align}
\lefteqn{G^{(\sigma,T)} (t,\mathbf{p})}\notag\\
& = \overline{U}^{(\sigma)} (T,t,\mathbf{p})G^{(\sigma,T=0)} (t,\mathbf{p})(\overline{U}^{(\sigma)})^{\rm T} (T,-t,-\mathbf{p}),\label{BT-generalmixed}
\end{align}
where
\begin{align}
\overline{U}^{(\sigma)}(T,t,\mathbf{p}) & = (n_{B}(E))^{\frac{1}{2}}\begin{pmatrix}
e^{\frac{\beta E}{2}} & P^{(\sigma)}(T,t,E)\\
P^{(\sigma)}(T,t,E) & e^{\frac{\beta E}{2}}
\end{pmatrix}\notag\\
& = (\overline{U}^{(\sigma)})^{\rm T}(T,t,\mathbf{p}). \label{mixedgeneral-BTmatrix}
\end{align}
Here $P^{(\sigma)}(T,t,\mathbf{p})$ denotes the function
\begin{align}
P^{(\sigma)}(T,t,\mathbf{p}) & = \theta(t) e^{-(\sigma-\frac{1}{2})\beta E} + \theta(-t) e^{(\sigma-\frac{1}{2})\beta E}\notag\\
& = P^{(1-\sigma)} (T,-t,\mathbf{p}),\label{P}
\end{align}
and we note that the matrix $\overline{U}^{(\sigma)} (T,t,\mathbf{p})$ in \eqref{mixedgeneral-BTmatrix} depends only on the magnitude $|\mathbf{p}|$.  Using this, it follows that
\begin{align}
(\overline{U}^{(\sigma)})^{\rm T} (T, -t,-\mathbf{p}) & = (\overline{U}^{(1-\sigma)})^{\rm T} (T,t,-\mathbf{p})\notag\\
& = \overline{U}^{(1-\sigma)} (T,t,\mathbf{p}).
\end{align}
We note that the factorizing matrix, $\overline{U}^{(\sigma)} (T,t,\mathbf{p})$, now depends on $t$ unlike in the closed time path (see \eqref{mixedCT-BTmatrix}). As a result, the finite temperature propagator, \eqref{BT-generalmixed}, in the mixed space can also be written as
\begin{align}
\lefteqn{G^{(\sigma,T)} (t,\mathbf{p})}\notag\\
& = \overline{U}^{(\sigma)} (T,t,\mathbf{p})G^{(\sigma,T=0)} (t,\mathbf{p})\overline{U}^{(1-\sigma)} (T,t,\mathbf{p}).\label{BT-generalmixed1}
\end{align}

\subsection{A possible scalar thermal operator representation}

As we have emphasized repeatedly, since the zero temperature propagator for $0<\sigma<1$ has only two diagonal elements, a (well behaved) scalar thermal operator, acting on the zero temperature propagator, cannot generate all four components of the thermal propagator. From the Bogoliubov transformation point of view, this can be seen in the following way. Using the matrix form in \eqref{mixedgeneral-BTmatrix} as well as \eqref{P}, we can work out \eqref{BT-generalmixed1} directly to give
\begin{widetext}
\begin{align}
\lefteqn{G^{(\sigma,T)} (t,\mathbf{p})
 = \overline{U}^{(\sigma)} (T,t,\mathbf{p})G^{(\sigma,T=0)} (t,\mathbf{p})\overline{U}^{(1-\sigma)} (T,t,\mathbf{p})}\notag\\
& = n_{B}(E)\begin{pmatrix}
e^{\beta E}G_{11}^{(\sigma,T=0)}(t,\mathbf{p}) + P^{(\sigma)}P^{(1-\sigma)}G_{22}^{(\sigma,T=0)}(t,\mathbf{p}) & e^{\frac{\beta E}{2}}(P^{(1-\sigma)}G_{11}^{(\sigma,T=0)}(t,\mathbf{p})+P^{(\sigma)}G_{22}^{(\sigma,T=0)}(t,\mathbf{p}))\\
\noalign{\vskip 2pt}%
e^{\frac{\beta E}{2}}(P^{(\sigma)}G_{11}^{(\sigma,T=0)}(t,\mathbf{p})+P^{(1-\sigma)}G_{22}^{(\sigma,T=0)}(t,\mathbf{p})) & e^{\beta E}G_{22}^{(\sigma,T=0)}(t,\mathbf{p}) + P^{(\sigma)}P^{(1-\sigma)}G_{11}^{(\sigma,T=0)}(t,\mathbf{p})
\end{pmatrix}\notag\\
&\neq {\cal O}^{(T)}_{\rm scalar} \begin{pmatrix}
G_{11}^{(\sigma,T=0)}(t,\mathbf{p}) & 0\\
0 & G_{22}^{(\sigma,T=0)}(t,\mathbf{p})
\end{pmatrix},
\end{align}
\end{widetext}
where ${\cal O}^{(T)}_{\rm scalar}$ is a scalar operator carrying the entire temperature dependence (not necessarily the same one as defined in \eqref{TO}).

Since a direct thermal operator representation as in \eqref{TO} is not possible for $0<\sigma<1$, let us ask if a modified thermal operator representation, starting from the Bogoliubov transformation, is possible which can be calculationally simple. To study this question, let us note that the Bogoliubov transformation matrix in \eqref{mixedgeneral-BTmatrix} is, in fact, factorizable as
\begin{equation}
\overline{U}^{(\sigma)} (T,t,\mathbf{p}) = \overline{V}^{(\sigma)} (T,t,\mathbf{p}) A^{(\sigma)} (T,t,\mathbf{p}),
\end{equation}
where
\begin{align}
\lefteqn{\overline{V}^{(\sigma)} (T,t,\mathbf{p})}\notag\\
& = (n_{B}(E))^{\frac{1}{2}}\begin{pmatrix} 
e^{\frac{\beta E}{2}} & \theta(-t)e^{(\sigma-\frac{1}{2})\beta E}\\
\theta(-t)e^{(\sigma-\frac{1}{2})\beta E} & e^{\frac{\beta E}{2}}
\end{pmatrix}\notag\\
& = (n_{B}(E))^{\frac{1}{2}}(e^{\frac{\beta E}{2}} \mathbbm{1} + \theta(-t) e^{(\sigma-\frac{1}{2})\beta E} \sigma_{1})\notag\\
& = (\overline{V}^{(\sigma)})^{\rm T} (T,t,\mathbf{p}),\label{Vbar}
\end{align}
and
\begin{align}
A^{(\sigma)}(T,t,\mathbf{p}) & = \begin{pmatrix}
1 & \theta(t) e^{-\sigma\beta E}\\
\theta(t) e^{-\sigma\beta E} & 1
\end{pmatrix}\notag\\
& = \mathbbm{1} + \theta(t) e^{-\sigma\beta E} \sigma_{1}\notag\\
& = (A^{(\sigma)})^{\rm T} (T,t,\mathbf{p}).\label{A}
\end{align}
Here $\sigma_{1}$ denotes the Pauli matrix. (Note that both $\overline{V}^{(\sigma)}$ and $A^{(\sigma)}$ depend only on the magnitude $|\mathbf{p}|$.)

As a result, we can write (see \eqref{BT-generalmixed})
\begin{align}
\lefteqn{G^{(\sigma,T)} (t,\mathbf{p})}\notag\\
& = \overline{U}^{(\sigma)} (T,t,\mathbf{p})G^{(\sigma,T=0)} (t,\mathbf{p})(\overline{U}^{(\sigma)})^{\rm T} (T,-t,-\mathbf{p})\notag\\
& = \overline{V}^{(\sigma)} (T,t,\mathbf{p}) \overline{G}^{(\sigma)}(t,\mathbf{p})\overline{V}^{(\sigma)}(T,-t,-\mathbf{p}),\label{factorized}
\end{align}
where
\begin{align}
\overline{G}^{(\sigma)} (t,\mathbf{p}) & = A^{(\sigma)}(T,t,\mathbf{p})G^{(\sigma,T=0)}(t,\mathbf{p}) A^{(\sigma)}(T,-t,-\mathbf{p})\notag\\
& = \begin{pmatrix}
G_{11}^{(\sigma,T=0)}(t,\mathbf{p}) & \frac{e^{i (t+i\sigma\beta)E}}{2E}\\
\noalign{\vskip 2pt}%
\frac{e^{-i (t-i\sigma\beta)E}}{2E} & G_{22}^{(\sigma,T=0)}(t,\mathbf{p})
\end{pmatrix}.\label{Gbar}
\end{align}
This matrix has all four components nontrivial and we can think of $\overline{G}^{(\sigma)}(t,\mathbf{p})$ as a generalization of $G^{(\sigma,T=0)}(t,\mathbf{p})$ with temperature dependent off-diagonal elements which vanish exponentially as $T\rightarrow 0$ ($\beta\rightarrow \infty$).

From \eqref{factorized}, we see that we can now write
\begin{align}
G^{(\sigma,T)}(t,\mathbf{p}) & = \overline{V}^{(\sigma)}(T,t,\mathbf{p}) \overline{V}^{(\sigma)}(T,-t,-\mathbf{p})\overline{G}^{(\sigma)}(t,\mathbf{p})\notag\\
&\quad + \overline{V}^{(\sigma)}(T,t,\mathbf{p})[\overline{G}^{(\sigma)}(t,\mathbf{p}), \overline{V}^{(\sigma)}(T,-t,-\mathbf{p})].\label{reorder}
\end{align}
Each term on the right hand side of \eqref{reorder} can be worked out easily using \eqref{Vbar}, the properties of the Pauli matrices and \eqref{Gbar} and lead to
\begin{align}
\lefteqn{G^{(\sigma,T)}(t,\mathbf{p})}\notag\\
& = ((1+n_{B}(E))\mathbbm{1} + n_{B}(E)e^{\sigma\beta E}\sigma_{1}) \overline{G}^{(\sigma)}(t,\mathbf{p})\notag\\
&\quad -n_{B}(E)(S(E) + e^{\sigma\beta E} \sigma_{1})\overline{G}^{(\sigma)}(t,\mathbf{p})\notag\\
& = (1+n_{B}(E)(1-S(E)))\overline{G}^{(\sigma)}(t,\mathbf{p})\notag\\
& = {\cal O}^{(T)}(E) \overline{G}^{(\sigma)}(t,\mathbf{p}).\label{TOR}
\end{align}
Here $S(E)$ is the reflection operator defined in \eqref{reflection} and ${\cal O}^{(T)}(E)$ is the same scalar thermal operator as for the closed time path in \eqref{TO}.

This analysis shows that while, for $0<\sigma<1$, there is no scalar thermal operator relating directly the finite temperature propagator to the zero temperature one, the same scalar thermal operator (as in the closed time path) relates the finite temperature propagator to a generalized form of the zero temperature propagator given in \eqref{Gbar}. This generalized form of the zero temperature propagator, in fact, coincides with the zero temperature propagator \eqref{mixedCT-zero} of the closed time path  when $\sigma=0$, thereby unifying the description for all paths (for $\sigma=1$, the propagator is the transpose of the one in closed time path as we have already discussed). Having this scalar thermal operator relation is indeed of great calculational help and leads to a better understanding of various phenomena.

Finally, we note here that the reflection operator $S(E)$ as well as the thermal operator ${\cal O}^{(T)}(E)$ in \eqref{TO}, \eqref{TO1} and \eqref{TOR} are independent of the time variable. As a result, the entire $t$ dependence of $G^{(\sigma,T)}(t,\mathbf{p})$ in \eqref{TOR} is contained in the generalized propagator $\overline{G}^{(\sigma)}(t,\mathbf{p})$ defined in \eqref{Gbar}. Consequently, if we were to inverse Fourier transform $G^{(\sigma,T)}(t,\mathbf{p})$ in \eqref{TOR} to the $p_{0}$ space, neither $S(E)$ nor ${\cal O}^{(T)}(E)$ would change and we can write
\begin{align}
\lefteqn{G^{(\sigma,T)}(p_{0},\mathbf{p}) = {\cal O}^{(T)}(E)\overline{G}^{(\sigma)}(p_{0},\mathbf{p})}\notag\\
&\qquad = (1 + n_{B}(E)(1-S(E))) \overline{G}^{(\sigma)}(p_{0},\mathbf{p}),\label{TORmomentum}
\end{align}
where the matrix components ($i,j=1,2$) are given by
\begin{equation}
\overline{G}^{(\sigma)}_{ij} (p_{0},\mathbf{p}) = \int_{-\infty}^{\infty} dt\,\overline{G}^{(\sigma)}_{ij} (t,\mathbf{p}).\label{components}
\end{equation}
Using \eqref{Gbar}, the momentum space components in \eqref{components} can be calculated (with regularization where needed) in a straight forward manner and have the forms 
\begin{align}
\overline{G}^{(\sigma)}_{11}(p_{0},\mathbf{p}) & = \frac{i}{p_{0}^{2} - E^{2} + i\epsilon},\notag\\
\overline{G}^{(\sigma)}_{12}(p_{0},\mathbf{p}) & = 2\pi \frac{e^{-\sigma\beta E}}{2E}  \delta (p_{0}+E),\notag\\
\overline{G}^{(\sigma)}_{21}(p_{0},\mathbf{p}) & = 2\pi \frac{e^{-\sigma\beta E}}{2E}  \delta (p_{0}-E),\notag\\
\overline{G}^{(\sigma)}_{22}(p_{0},\mathbf{p}) & = - \frac{i}{p_{0}^{2} - E^{2} - i\epsilon}.\label{components1}
\end{align}
We note that these components of $\overline{G}^{(\sigma)}_{ij}(p_{0},\mathbf{p})$ do indeed coincide with \eqref{CT-zero} when $\sigma=0$. Furthermore, substituting these components into \eqref{TORmomentum}, we easily verify that  this leads to \eqref{generalpath}. This shows that a scalar thermal operator relation holds in momentum space as well. However, as we have pointed out earlier, calculations are much simpler in the mixed space.

\section{Conclusion} 

Calculations of thermal amplitudes are, in general, simpler in the mixed space ($t,\mathbf{p}$) \cite{das}, compared to the zero temperature ones, although still quite involved since the thermal propagators have nontrivial forms (both in the imaginary time and the real time formalisms). We note that the interaction vertices, in a thermal field theory, have no temperature dependence although in the real time formalism they have a $2\times 2$ matrix structure (proportional to $\sigma_{3}$). 

In the imaginary time formalism, although the degrees of freedom do not double, the Euclidean time interval ($0\leq\tau\leq\beta$) is finite at finite temperature as opposed to zero temperature where $-\infty<\tau<\infty$. Therefore, a direct relation between finite temperature graphs and the corresponding zero temperature ones seems unlikely. Nonetheless, the existence of a multiplicative (scalar) thermal operator relating the finite temperature and the zero temperature propagators in the mixed space leads to a simple relation between Feynman amplitudes graph by graph \cite{brandt,brandt1}.

In the real time formalisms, the number of degrees of freedom doubles and the propagators become $2\times 2$ matrices (both at zero and finite temperatures). The finite temperature propagators are related to the zero temperature ones through a temperature dependent $2\times 2$ Bogoliubov transformation matrix in the momentum space. Since the interaction vertices do not carry temperature dependence, this implies that in the real time formalisms there is, in principle, a direct matrix relation between a thermal graph and the corresponding zero temperature graph in momentum space. This is, however, not very useful calculationally since keeping track of the matrix indices becomes tedious in a complicated Feynman graph.

On the other hand, it was shown \cite{brandt,brandt1} that, in the closed time path ($\sigma=0$), the finite temperature (matrix) propagator in the mixed space is, in fact, related to the zero temperature (matrix) propagator by a multiplicative scalar thermal operator. This also turns out to be the case for $\sigma=1$. Therefore, in these two cases, a thermal Feynman graph is simply related to the corresponding zero temperature one by a product of the (multiplicative) scalar operators carrying the entire temperature dependence. This is not only very useful calculationally, but is of considerable help in all order proofs of certain results at finite temperature \cite{niegawa,brandt2,brandt3,brandt4,frenkel,frenkel1}. It had also been shown \cite{brandt} that such a factorization of the thermal propagator (and, therefore, of thermal Feynman graphs) in terms of a multiplicative scalar thermal operator is not possible for other real time formalisms corresponding to $0<\sigma<1$.

In this paper, we have shown how the scalar thermal operator, in the case of closed time path ($\sigma=0$) as well as $\sigma=1$, in the mixed space, arises from the momentum space $2\times 2$ Bogoliubov transformation matrix in a thermal field theory. Starting with the factorization in terms of the Bogoliubov transformation matrix in momentum space, we have also shown why a scalar thermal operator representation fails in the mixed space when $0<\sigma<1$. On the other hand, a systematic analysis following from the Bogoliubov transformation matrix shows that a scalar thermal operator acting on a limiting form of the zero temperature propagator can give rise to the thermal propagator in this case ($0<\sigma<1$) which can be calculationally useful. Furthermore, this limiting form of the zero temperature propagator coincides with that of the closed time path when $\sigma=0$ unifying the thermal operator representation for all paths. This analysis also shows the existence of a scalar thermal operator relation in the momentum space.
\medskip

\noindent {\bf Acknowledgment}
\medskip

A. D. (Atri Deshamukhya) would like to thank IOP, Bhubaneswar, for hospitality where part of this work was done. S. P. acknowledges financial support received through his J. C. Bose fellowship.

\appendix

\section{Derivation of the factorization matrix in the mixed space from the Bogoliubov transformation matrix}

In this appendix, we give a derivation of how the factorizing matrix $\overline{U}^{(\sigma)} (T,t,\mathbf{p})$ in the mixed space (see \eqref{BT-generalmixed}-\eqref{P}) arises from the more fundamental Bogoliubov transformation matrix $U^{(\sigma)} (T,p)$ in the momentum space defined in \eqref{BT-generalp} and \eqref{pgeneral-BTmatrix}. Denoting the matrix elements of $U^{(\sigma)}(T,p)$ in \eqref{pgeneral-BTmatrix} as $U^{(\sigma)}_{ij}, i,j=1,2$ for simplicity (without writing explicitly the dependence on $(T,p)$), the matrix elements of the finite temperature propagtor in \eqref{BT-generalp1} can be written as (recall that the propagator $G^{(\sigma,T=0)}(p)$ is a diagonal matrix)
\begin{align}
G^{(\sigma,T)}_{11}(p) & = U^{(\sigma)}_{11}G_{11}^{(\sigma,T=0)}(p)U^{(\sigma)}_{11} + U^{(\sigma)}_{12}G_{22}^{(\sigma,T=0)}(p)U^{(\sigma)}_{21},\notag\\
G^{(\sigma,T)}_{12}(p) & = U^{(\sigma)}_{11}G_{11}^{(\sigma,T=0)}(p)U^{(\sigma)}_{12} + U^{(\sigma)}_{12}G_{22}^{(\sigma,T=0)}(p)U^{(\sigma)}_{22},\notag\\
G^{(\sigma,T)}_{21}(p) & = U^{(\sigma)}_{21}G_{11}^{(\sigma,T=0)}(p)U^{(\sigma)}_{11} + U^{(\sigma)}_{22}G_{22}^{(\sigma,T=0)}(p)U^{(\sigma)}_{21},\notag\\
G^{(\sigma,T)}_{22}(p) & = U^{(\sigma)}_{21}G_{11}^{(\sigma,T=0)}(p)U^{(\sigma)}_{12} + U^{(\sigma)}_{22}G_{22}^{(\sigma,T=0)}(p)U^{(\sigma)}_{22}.\label{BT-generalp2}
\end{align} 
We note from \eqref{pgeneral-BTmatrix} that the matrix elements,  $U^{(\sigma)}_{ij}$, of the Bogoliubov transformation matrix are functions only of $p^{0}$ (and, of course, $T$). Furthermore, the diagonal components of the zero temperature propagator, given in \eqref{general-zero}, can also be written as (from the point of view of taking the Fourier transform)
\begin{align}
G_{11}^{(\sigma,T=0)}(p) & = \frac{i}{(p_{0}-E+i\epsilon)(p_{0}+E-i\epsilon)},\notag\\
G_{22}^{(\sigma,T=0)}(p) & = -\frac{i}{(p_{0}-E-i\epsilon)(p_{0}+E+i\epsilon)},\label{rewriteGzero}
\end{align}
with $E$ defined in \eqref{E}. The two components of the propagator in \eqref{rewriteGzero} have poles at $p_{0}=\pm(E-i\epsilon)$ and $p_{0}=\pm(E+i\epsilon)$ respectively.

The Fourier transforms of the two diagonal components in \eqref{rewriteGzero} with respect to $p_{0}$ leads to the components, $G_{11}^{(\sigma,T=0)} (t, \mathbf{p})$ and $G_{22}^{(\sigma,T=0)} (t,\mathbf{p})$, of the zero temperature propagator in the mixed space given in \eqref{mixedgeneral-zero}. We note next, from \eqref{rewriteGzero}, that the Fourier transform of a product of functions together with the diagonal components can be obtained simply as
\begin{align}
\lefteqn{\int \frac{dp_{0}}{2\pi}\,e^{-ip_{0}t}\,f(p_{0}) G_{11}^{(\sigma,T=0)}(p) g(p_{0})}\notag\\
& = \frac{1}{2E}\left(\theta(t)f(E)e^{-iEt}g(E)+\theta(-t)f(-E)e^{iEt}g(-E)\right)\notag\\
& = (\theta(t)f(E)+\theta(-t)f(-E))\frac{1}{2E}\left(\theta(t)e^{-iEt}+\theta(-t)e^{iEt}\right)\notag\\
&\qquad\times(\theta(t)g(E)+\theta(-t)g(-E))\notag\\
& = f(t,E)G_{11}^{(\sigma,T=0)}(t,\mathbf{p})g(t,E),\label{identity1}
\end{align}
and similarly, 
\begin{align}
\lefteqn{\int \frac{dp_{0}}{2\pi}\,e^{-ip_{0}t}\,f(p_{0}) G_{22}^{(\sigma,T=0)}(p) g(p_{0})}\notag\\
&\quad = f(t,-E) G_{22}^{(\sigma,T=0)}(t,\mathbf{p})g(t,-E).\label{identity2}
\end{align}
Here we have identified
\begin{align}
f(t,E) & = \theta(t)f(E)+\theta(-t)f(-E),\notag\\ 
g(t,E) &= \theta(t)g(E)+\theta(-t)g(-E).\label{fg}
\end{align}

If we write the finite temperature propagator in the mixed space, $G^{(\sigma,T)}(t,\mathbf{p})$, in the factorized form $\overline{U}^{(\sigma)}G^{(\sigma,T=0)}(t,\mathbf{p})\widetilde{U}^{(\sigma)}$ (see, for example, \eqref{BT-generalmixed1}, namely, the matrices on the left and right do not have to be the same unlike in the case of the closed time path \eqref{matrixfactorization}), the components take the form
\begin{widetext}
\begin{align}
G^{(\sigma,T)}_{11}(t,\mathbf{p}) & = \overline{U}^{(\sigma)}_{11}G_{11}^{(\sigma,T=0)}(t,\mathbf{p})\widetilde{U}^{(\sigma)}_{11} + \overline{U}^{(\sigma)}_{12}G_{22}^{(\sigma,T=0)}(t,\mathbf{p})\widetilde{U}^{(\sigma)}_{21},\notag\\
G^{(\sigma,T)}_{12}(t,\mathbf{p}) & = \overline{U}^{(\sigma)}_{11}G_{11}^{(\sigma,T=0)}(t,\mathbf{p})\widetilde{U}^{(\sigma)}_{12} + \overline{U}^{(\sigma)}_{12}G_{22}^{(\sigma,T=0)}(t,\mathbf{p})\widetilde{U}^{(\sigma)}_{22},\notag\\
G^{(\sigma,T)}_{21}(t,\mathbf{p}) & = \overline{U}^{(\sigma)}_{21}G_{11}^{(\sigma,T=0)}(t,\mathbf{p})\widetilde{U}^{(\sigma)}_{11} + \overline{U}^{(\sigma)}_{22}G_{22}^{(\sigma,T=0)}(t,\mathbf{p})\widetilde{U}^{(\sigma)}_{21},\notag\\
G^{(\sigma,T)}_{22}(t,\mathbf{p}) & = \overline{U}^{(\sigma)}_{21}G_{11}^{(\sigma,T=0)}(t,\mathbf{p})\widetilde{U}^{(\sigma)}_{12} + \overline{U}^{(\sigma)}_{22}G_{22}^{(\sigma,T=0)}(t,\mathbf{p})\widetilde{U}^{(\sigma)}_{22}.\label{BT-generalmixed2}
\end{align} 
\end{widetext} 
Each component of the finite temperature propagator in mixed space is, of course, the Fourier transform of the corresponding component in the momentum space, namely,
\begin{equation}
G_{ij}^{(\sigma,T)}(t,\mathbf{p}) = \int \frac{dp_{0}}{2\pi}\,e^{-ip_{0}t}\,G_{ij}^{(\sigma,T)}(p).
\end{equation}
Therefore, using the relations in \eqref{BT-generalp2} and \eqref{identity1}-\eqref{fg}, we can determine
\begin{align}
\overline{U}^{(\sigma)}_{11} & = \theta(t) U^{(\sigma)}_{11}(E) + \theta(-t) U^{(\sigma)}_{11}(-E)\notag\\
& = (n_{B}(E))^{\frac{1}{2}}e^{\frac{\beta E}{2}},\notag\\
\overline{U}^{(\sigma)}_{12} & = \theta(t)U^{(\sigma)}_{12}(-E)+\theta(-t)U^{(\sigma)}_{12}(E)\notag\\
& = (n_{B}(E))^{\frac{1}{2}}(\theta(t)e^{-(\sigma-\frac{1}{2})\beta E} + \theta(-t)e^{(\sigma-\frac{1}{2})\beta E}),\notag\\
\overline{U}^{(\sigma)}_{21} & = \theta(t) U^{(\sigma)}_{21}(E) + \theta(-t) U^{(\sigma)}_{21}(-E)\notag\\
&= (n_{B}(E))^{\frac{1}{2}}(\theta(t)e^{-(\sigma-\frac{1}{2})\beta E} + \theta(-t)e^{(\sigma-\frac{1}{2})\beta E})\notag\\
& = \overline{U}^{(\sigma)}_{12},\notag\\
\overline{U}^{(\sigma)}_{22} &= \theta(t) U^{(\sigma)}_{22}(-E) + \theta(-t) U^{(\sigma)}_{22}(E)\notag\\
& = (n_{B}(E))^{\frac{1}{2}} e^{\frac{\beta E}{2}} = \overline{U}^{(\sigma)}_{11}.\label{FT1}
\end{align}
Similarly, we can show that
\begin{align}
\widetilde{U}^{(\sigma)}_{11} & = (n_{B}(E))^{\frac{1}{2}}e^{\frac{\beta E}{2}},\notag\\
\widetilde{U}^{(\sigma)}_{12} & = (n_{B}(E))^{\frac{1}{2}}(\theta(t)e^{(\sigma-\frac{1}{2})\beta E} + \theta(-t)e^{-(\sigma-\frac{1}{2})\beta E}),\notag\\
\widetilde{U}^{(\sigma)}_{21} & = (n_{B}(E))^{\frac{1}{2}}(\theta(t)e^{(\sigma-\frac{1}{2})\beta E} + \theta(-t)e^{-(\sigma-\frac{1}{2})\beta E}),\notag\\
& = \widetilde{U}^{(\sigma)}_{12},\notag\\
\widetilde{U}^{(\sigma)}_{22} & = (n_{B}(E))^{\frac{1}{2}}e^{\frac{\beta E}{2}} = \widetilde{U}^{(\sigma)}_{11}.\label{FT2}
\end{align}
Equation \eqref{FT1} shows that the matrix $\overline{U}^{(\sigma)}$ in \eqref{mixedgeneral-BTmatrix} does indeed correspond to the matrix obtained from a Fourier transformation of \eqref{BT-generalp1} to the mixed space while \eqref{FT2} identifies
$\widetilde{U}^{(\sigma)} = \overline{U}^{(1-\sigma)}$,
so that the factorization in \eqref{BT-generalmixed1} follows from the Fourier transform of the Bogoliubov transformation relation \eqref{BT-generalp1}.

\end{document}